# Radio Frequency Propagation Model and Fading of Wireless Signal at 2.4 GHz in Underground Coal Mine

Ashutosh Patri[1], Devidas S. Nimaje[2]

Department of Mining Engineering, National Institute of Technology Rourkela

Rourkela, Odisha, India

E-mail address: ashutoshpatri@gmail.com[1], dsnimaje@nitrkl.ac.in[2]

## Abstract

Deployment of wireless sensor networks and wireless communication systems have become indispensable for better real-time data acquisition from ground monitoring devices, gas sensors, and equipment used in underground mines as well as in locating the miners, since conventional methods like use of wireline communication are rendered ineffective in the event of mine hazards such as roof-falls, fire hazard etc. Before implementation of any wireless system, the variable path loss indices for different work place should be determined; this helps in better signal reception and sensor-node localisation. This also improves the method by which miner carrying the wireless device is tracked. This paper proposes a novel method for parameter determination of a suitable radio propagation model with the help of results of a practical experiment carried out in an underground coal mine of Southern India. The path loss indices along with other essential parameters for accurate localisation have been determined using XBee module and ZigBee protocol at 2.4 GHz frequency.

## 1. Introduction

There have been several advancements in mining industry in last three decades, which have primarily focused on improvements in heavy machineries, support systems and safety equipment. Recently the focus has shifted towards development of communication systems for better safety and connectivity. In this context, Wireless Sensor Network (WSN) owing to its efficiency, speed and applicability in emergency conditions has come out on top [1-3]. The need of the hour is to achieve a reliable wireless system in the harsh underground mine environment [4], in which radio propagation models are playing a vital role.

Recent studies have considered the underground mine as a hybrid case of regular and harsh environment and shown that the signal propagation models and critical parameters of wireless channel propagation for indoor environment is similar to underground mine scenario at 900 MHz, indicating that the wireless motes used in the indoor environment can be modified for use in mines [5, 6]. Zhang et al. experimented at 900 MHz with two different scenarios namely passageway and mining zone of longwall coal mine in order to evaluate the additional losses due to passage-way curvature and presence of coal mining equipment and subsequently modified the wave guide propagation model. The hybrid tunnel propagation model developed by them uses both free space propagation model and a modified waveguide propagation model to describe the propagation characteristics [7]. Some simulation tools have also been developed for path loss calculation and propagation modelling by taking into account the effects of barriers. The simulations were carried out by varying the frequency with standard tunnel dimension, shape and material properties; on comparison with an actual scenario, it was proved that the path loss is mostly dependent on tunnel dimension and signal frequency [8].

With advancement in Micro-Electro Mechanical Systems (MEMS), nowadays transceivers working at 2.4GHz are available at a reasonable price [9]. The better performance of such transceivers in localisation within a small range is owing to high-directivity antenna and very high operational frequency resulting in presence of relatively less additional noise. Liu et al. studied the transmission performance of WSN near mine-working face at 2.4 GHz frequency and incorporated all the electromagnetic properties in their theoretical model and compared it with experimental results. The effective transmission distance was studied for IEEE 802.15.4 known as ZigBee protocol [10, 11].

In this paper, the Radio Frequency (RF) propagation model has been prepared and the path loss of wireless signal at 2.4 GHz has been experimentally derived for the GDK 10A incline, a longwall underground mine of Singareni Collieries Company Limited (SCCL). Before implementing WSN, the path loss index and other parameters should be calculated to perform better localisation, base-station placement and optimisation, improvement in receiver design and combating the fading of signal [12]. The reception distance was determined by knowing the path loss of signal which decides the energy loss factor. The repeaters should be placed accordingly and their amplification factors should be set to different values to achieve a high efficiency wireless communication system for different environment. The performance of ZigBee protocol using XBee module was experimentally studied for the aforesaid mine.

## 2. Radio frequency propagation models

A wireless propagation model can be defined as a mathematical expression or an algorithm for predicting radio characteristics of a particular type of environment. There are two types of wireless propagation model i.e. deterministic model and empirical model [12, 13]. The deterministic model does not fit into the real environment properly; however for low frequency waves, the results produced by the deterministic model are approximately equal to the actual result, with a very low rounding-error. Since the operating range is very less, elements present in the surroundings have a significant effect on propagation in the high frequency channel while variations due to environmental effects are largely insignificant in low frequency channel. The propagation models for wireless network are categorised into three types i.e. free space propagation model, two-way ground model and log normal model [12]. These models are deterministic with the exception of the log-normal model, which is empirical.

### *2.1. Free space propagation model*

It is a simplified model which assumes line of sight communication between the transmitter-receiver pair and there is no obstruction present between them. The mathematical representation of the model can be written as

$$P_r(d) = C_T \left(\frac{P_t}{d^2}\right) \tag{1}$$

Where, $P_r$ and $P_t$ represent the power received and power transmitted respectively. $C_T$ is the constant depending on the transceiver and $d$ is the distance between the transmitter-receiver pair.

### *2.2. Two-ray ground model*

This model is obtained by modifying the above model after taking into account the effect of reflection of signals. It is also assumed that both direct and reflected rays are used for communication. In this model the distance between the transmitter-receiver pair is much greater than the height of their individual heights and it can be represented as

$$P_r(d) = C_t \left(\frac{P_t}{d^4}\right) \tag{2}$$

Where, $C_t$ is the constant representing transceiver characteristic in the two-ray ground model.

### *2.3. Log-distance model*

It is an analytical and empirical model which can be mathematically represented as

$$P_r(d) \propto \left(\frac{P_t}{d^n}\right) \tag{3}$$

Where, $\eta$ represents the path loss factor or distance power gradient.

The experimental results vary from derived ones and hence for hostile environments like underground mines, the required model can be derived from shadow fading phenomenon.

At high frequency, power loss is different for different locations owing to obstructions present in the path between two communicating devices. In Figure 1, a typical illustration of this fact is given where the dotted circle shows the ideal boundary of operation for an omnidirectional antenna placed at the centre, but the bold line shows the actual boundary of operation with a minimum and maximum range of $R_1$ and $R_2$ respectively due to presence of various obstructions. For this purpose empirical model is chosen over the deterministic model to predict or calculate power received at a particular distance from the transmitter [14].

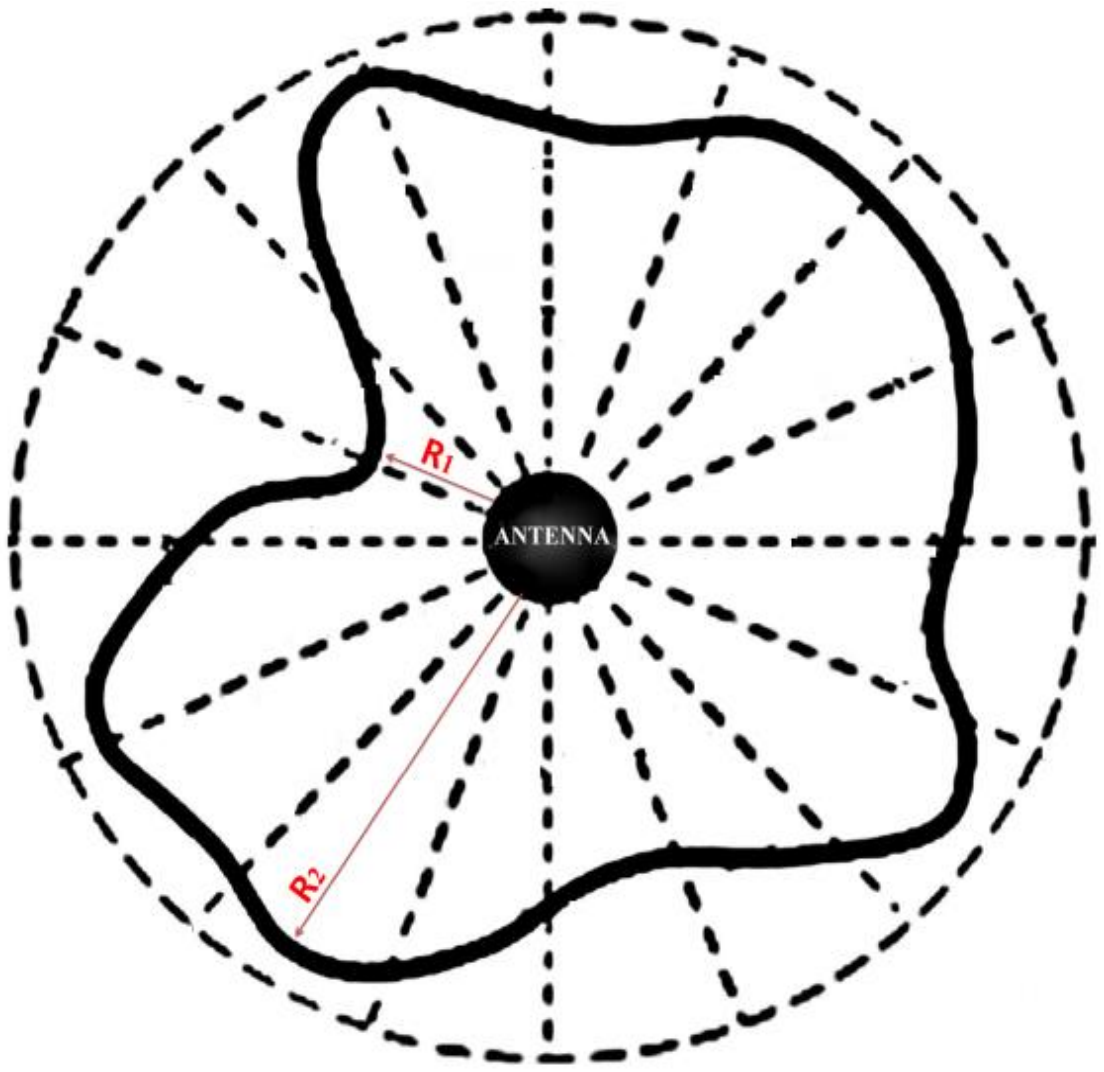

Fig.1: Variation in operation range due to fading of signal radiated from the omnidirectional antenna

Moreover the power loss can be subdivided into two parts on basis of fluctuation around the average path loss, i.e. Multi-path fading and Shadow fading. In case of Multi-path fading, the transmitted signal reaches the receiver through two or more paths causing both constructive and destructive interferences near the receiver which in turn leads to phase shifting and addition of noise. So, in a dynamic environment, where both the transmitter and receiver are stationary, the Received Signal Strength (RSS value) varies randomly due to the movement of objects and small changes in the environment. The long-term average of RSS values represents the effect of shadow fading of signal that is caused by the presence of a constant barrier present between the transceivers [14].

Although Time of Arrival (TOA), Angle of Arrival (AOA) and Time Difference of Arrival (TDOA) provide higher accuracy in most cases, they fail in a harsh mining environment [15, 16]. Therefore, Received Signal Strength Index (RSSI) based model for localisation has been developed. This low-cost RSSI based localisation provides less communication overhead with lower implementation complexity.

The distance or range of signal could be calculated accordingly to the loss factor of the environment from the RSSI based equations as given in (4) and (7).

### *2.4. Shadow fading model and proposed scheme for parameter determination*

The Log distance model can be represented more accurately by introducing a Gaussian distribution variable to represent the fading or fluctuation of received signal strength. The modified model is called Log-normal Shadowing model and it is most appropriate for wireless sensor networks since it is all inclusive in nature and can be easily configured according to the target environment [17]. The mathematical equation for the above relation can be defined as

$$PL(d)(\mathrm{dB}) = \overline{PL}(d_0)(\mathrm{dB}) + 10\eta \log_{10}\left(\frac{d}{d_0}\right) + \psi(\mathrm{dB}) \tag{4}$$

Where,

$$PL(\mathrm{dB}) = 10 \log_{10}\left[\frac{P_t}{P_r}\right] \tag{5}$$

and $d_0$ is the near earth reference distance. The random variable $\psi$ is the Zero-mean Gaussian random noise whose probability distribution fiction is given by

$$p(\psi_{\mathrm{dB}}) = \frac{1}{\sqrt{2\pi}\sigma_{\psi_{\mathrm{dB}}}} \exp\left[\frac{-\left(\psi_{\mathrm{dB}} - \mu_{\psi_{\mathrm{dB}}}\right)^2}{2\sigma^2{}_{\psi_{\mathrm{dB}}}}\right] \tag{6}$$

The value of $\eta$ depends on the surrounding or propagation environment as per equation (4). The distance $d_0$ is taken to be one meter for simplicity of calculation and it can also be represented in the terms of received power or RSSI as

$$\left[\frac{\overline{P_r}(d_0)}{P_r(d)}\right] = \left[\frac{d}{d_0}\right]^{\eta} + \psi \tag{7}$$

In equation (4), there are two unknown terms i.e. $\eta$ and $\psi$ which should be determined from experiments. The linear regression analysis for the data set with distance and received power as attributes gives the $\eta$ value, which can be further used for that particular place with unknown distance and known received power to localise a wireless node.

In equation (4), the $\mathrm{Var}(\psi) = \sigma^2$ and $\mathrm{E}(\psi) = 0$, so it can be mathematically proven that $\mathrm{Var}(\sigma\psi_1) = \sigma^2$ and $\mathrm{E}(\sigma\psi_1) = 0$. This relation shows that the $\psi$ function has the same distribution as $\psi_1$, where $\psi_1$ represents Zero-mean Gaussian distribution with unit variance. Equation (4) can be modified as

$$PL(d)(\mathrm{dB}) = \overline{PL}(d_0)(\mathrm{dB}) + 10\eta\log_{10}\left(\frac{d}{d_0}\right) + \sigma\psi_1\ (\mathrm{dB}) \tag{8}$$

Assuming maximum error with 95% confidence interval, the $\sigma\psi_1$ value can be replaced by 1.96 $\sigma$, which gives

$$\sigma_{i(exp)} = Y_i = \left(\frac{PL(d_i)}{1.96}\right) - \left[\left(\frac{\overline{PL}(d_0)}{1.96}\right) + \left(\frac{10\eta\log_{10} d_i}{1.96}\right)\right] \tag{9}$$

But, from the experiment carried out in the coal mine, the observational analysis shows that the standard deviation varies as a function of distance and on the basis of huge amount of experimental evidence, we claim it to be a forth degree polynomial function

$$\sigma_{i(obs)} = \mathrm{a}{d_i}^4 + \mathrm{b}{d_i}^3 + \mathrm{c}{d_i}^2 + \mathrm{e}d_i + \mathrm{f} \tag{10}$$

Now the observational error $\varepsilon$ can be defined as the difference of these two terms i.e., experimental and observational $\sigma$.

$$\varepsilon = \sigma_{i(exp)} - \sigma_{i(obs)} \tag{11}$$

For avoiding negative error and for solving this, the objective function $\epsilon$ can be written as

$$\epsilon = \sum_{i=1}^{n}\left[Y_i - \left(\mathrm{a}{d_i}^4 + \mathrm{b}{d_i}^3 + \mathrm{c}{d_i}^2 + \mathrm{e}d_i + \mathrm{f}\right)\right]^2 \tag{12}$$

To obtain the values of the coefficients of the polynomial, i.e. a, b, c, e, and f, partial derivative method is adopted and it can be mathematically represented as the following set of equations

$$\frac{\partial\epsilon}{\partial a} = -2\sum_{i=1}^{n}\left[\left(Y_i - \left(\mathrm{a}{d_i}^4 + \mathrm{b}{d_i}^3 + \mathrm{c}{d_i}^2 + \mathrm{e}d_i + \mathrm{f}\right)\right){d_i}^4\right] = 0 \tag{13.1}$$

$$\frac{\partial\epsilon}{\partial b} = -2\sum_{i=1}^{n}\left[\left(Y_i - \left(\mathrm{a}{d_i}^4 + \mathrm{b}{d_i}^3 + \mathrm{c}{d_i}^2 + \mathrm{e}d_i + \mathrm{f}\right)\right){d_i}^3\right] = 0 \tag{13.2}$$

$$\frac{\partial\epsilon}{\partial c} = -2\sum_{i=1}^{n}\left[\left(Y_i - \left(\mathrm{a}{d_i}^4 + \mathrm{b}{d_i}^3 + \mathrm{c}{d_i}^2 + \mathrm{e}d_i + \mathrm{f}\right)\right){d_i}^2\right] = 0 \tag{13.3}$$

$$\frac{\partial\epsilon}{\partial e} = -2\sum_{i=1}^{n}\left[\left(Y_i - \left(\mathrm{a}{d_i}^4 + \mathrm{b}{d_i}^3 + \mathrm{c}{d_i}^2 + \mathrm{e}d_i + \mathrm{f}\right)\right)d_i\right] = 0 \tag{13.4}$$

$$\frac{\partial \epsilon}{\partial f} = -2\sum_{i=1}^{n}\left(Y_i - \left(\mathrm{a}d_i^{\,4} + \mathrm{b}d_i^{\,3} + \mathrm{c}d_i^{\,2} + \mathrm{e}d_i + \mathrm{f}\right)\right) = 0 \qquad (13.5)$$

The above set of equations can be solved in the matrix form, to obtain the coefficients

$$\begin{bmatrix} \sum_{i=1}^{n} d_i^{\,8} & \sum_{i=1}^{n} d_i^{\,7} & \sum_{i=1}^{n} d_i^{\,6} & \sum_{i=1}^{n} d_i^{\,5} & \sum_{i=1}^{n} d_i^{\,4} \\ \sum_{i=1}^{n} d_i^{\,7} & \sum_{i=1}^{n} d_i^{\,6} & \sum_{i=1}^{n} d_i^{\,5} & \sum_{i=1}^{n} d_i^{\,4} & \sum_{i=1}^{n} d_i^{\,3} \\ \sum_{i=1}^{n} d_i^{\,6} & \sum_{i=1}^{n} d_i^{\,5} & \sum_{i=1}^{n} d_i^{\,4} & \sum_{i=1}^{n} d_i^{\,3} & \sum_{i=1}^{n} d_i^{\,2} \\ \sum_{i=1}^{n} d_i^{\,5} & \sum_{i=1}^{n} d_i^{\,4} & \sum_{i=1}^{n} d_i^{\,3} & \sum_{i=1}^{n} d_i^{\,2} & \sum_{i=1}^{n} d_i \\ \sum_{i=1}^{n} d_i^{\,4} & \sum_{i=1}^{n} d_i^{\,3} & \sum_{i=1}^{n} d_i^{\,2} & \sum_{i=1}^{n} d_i & n \end{bmatrix} \begin{bmatrix} \mathrm{a} \\ \mathrm{b} \\ \mathrm{c} \\ \mathrm{e} \\ \mathrm{f} \end{bmatrix} = \begin{bmatrix} \sum_{i=1}^{n} Y_i d_i^{\,4} \\ \sum_{i=1}^{n} Y_i d_i^{\,3} \\ \sum_{i=1}^{n} Y_i d_i^{\,2} \\ \sum_{i=1}^{n} Y_i d_i \\ \sum_{i=1}^{n} Y_i \end{bmatrix} \qquad (14)$$

By knowing the coefficients and path loss index for a particular place the standard deviation and the power loss due to fading can be calculated for new set of data with known RSSI and unknown distance, for accurate localisation.

## 3. Mining conditions of GDK 10A mine

GDK 10A incline of SCCL is situated at Ramagundam in Telengana, India in the Godavari valley coalfield. In Figure 2 schematic layout of a longwall mine is provided. The minimum and maximum depths of Seam-1, where experiments were carried out, are 175m and 310m respectively with a seam-thickness of 6.5m.

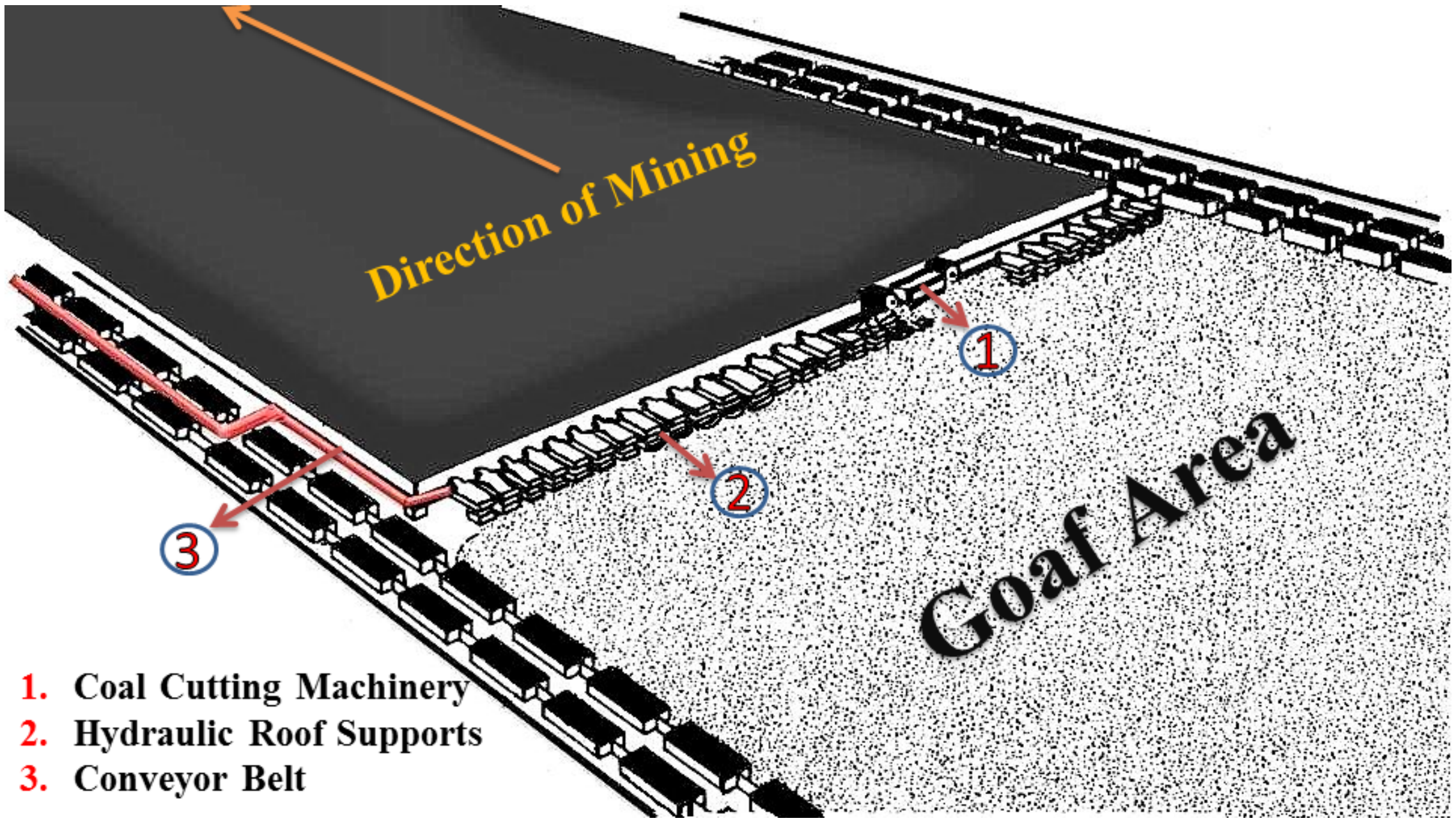

Fig.2: Schematic layout of Longwall mining method

The surface area is flat with undulating terrain having a gentle slope towards north-east and south. The coal seam is approached by driving two tunnels having a length of 450m and 500m at gradient of 1 in 4.5 and 1 in 5 for haulage and man-way respectively. The mine floor is mainly grey sand stone and has a coal roof with a clay band of 0.30m.

The length and width of Longwall-face is around 150m and 1km respectively with an average depth of 350m from surface. Anderson-made Double Ended Ranging Drum shearer having a diameter of 1.83m and a web width of 0.85m is used to mine coal. Caterpillar-made Independent Front Suspension based hydraulic powered roof supports are provided with 101 PMC-R controlled hydraulic-chocks. Anderson-made bridge-type stage loaders are used in gate-road to transport the coal from Armoured Face Conveyor (AFC) to belt conveyor. The 260m long DBT-made AFC is used in the face with a pan size of $232 \times 844 \times 1500\ mm^3$ and deck plate thickness of 35mm at an average chain speed of 1 m $s^{-1}$.

Both the head and tail gate-roads are parallel driven through seam-1. The gate-road wall surface is rough and water percolates from the strata and the gate-roads. The gate-road bearing the belt conveyor system has an average height and width of 3.6 and 4.2m respectively. The conveyor belt, supported by a steel structure, is at a height of 1.32m to 1.4m from the floor and carries an average lump size of $200 \times 200 \times 200\ mm^3$. The belt has a width of 0.8m to 1.2m mainly made of rubber. The roof supports are generally wire mesh type with bolts and girders. The material property, dimension and other features of the equipment described have a lot of influence on signal propagation along with mine dimension, rock property, slope and other geo-mining conditions.

## 4. Experimental set-up and procedure

### *4.1. Instruments and setup*

A pair of XBee series-1 modules, one being used as a transmitter and the other as a receiver, which implement ZigBee protocol, each capable of transmission or reception, were used for wireless communication at 2.4 GHz. The specification of the XBee module is given in Table 1. Each of the XBee modules has a mounted rubber duck wire antenna and is configured by setting the preferred data rate, modulation technique, lapse rate between packets and other parameters using X-CTU software by mounting the modules on the XBee USB adapter (which has an on-board 3.3V low drop voltage regulator and Light Emitting Diode (LED) indicators for RSSI, Associate and Power) and then connected to a computer's Universal Serial Bus (USB) port through a FT232 USB to serial converter. There are two modes of operation for XBee module; in Transparent Data Mode (AT) the signal coming to the Data IN (DIN) pin is directly sent to the receivers, while in Application Programming Interface Mode (API) (which was used in this study), the data is sent in the form of packets that include the receiver address along with a feedback for the delivered packets, payload information and various parameter settings to increase the reliability of the network and to send the signal safely over the wireless network [18].

Table 1. Specifications of XBee module

| **Parameter** | **Property** |
| --- | --- |
| Raw Data Rate | 2.4 GHz: 250 kbps (ISM band) |
| Maximum Range | Indoor: 30m; Outdoor (Line of Sight): 100m |
| Receiver sensitivity | -92 dBm (1% Packet Error Rate) |
| Channels | 16 channels |
| Addressing | Short 8 bit or 64 bit IEEE |
| Temperature | -40 to +85 deg. Celsius |
| Channel access | CSMA-CA (Carrier Sense Multi Access- Collision Avoidance) |

This module also supports Universal Asynchronous Receiver/Transmitter (UART) Interface which is beneficial for clock setting and connecting it to a microcontroller. The ATMEL Atmega-32 Microcontroller (14.7456MHz crystal) development board was used which has a compatible UART serial communication integrated circuit along with Electrically Erasable Programmable Read Only Memory (EEPROM), Static Random Access Memory (SRAM) and in-system self-programmable flash memory of 1024, 2k and 32k bytes respectively. It has an in-built reverse polarity protection and the 7805 voltage regulator has a heat sink for continuous dissipation to supply 1amp current constantly without over-heating. The Request to Send (RTS) and Clear to Send (CTS) module pins can be used to provide flow control. CTS flow control provides an indication to the host to stop sending serial data to the module. RTS flow control allows the host to signal the module not to send data in the serial-transmit buffer through the UART. Data in the serial-transmit buffer will not be sent out through the Data OUT (DOUT) pin as long as RTS is de-asserted or set high. The UART connections for the transmitter and receiver module are shown in Figure 3. The module operates in a low voltage range of 2.8-3.4 volt, but for the whole setup, a pair of 12V- 1.3Ah DC battery of lead-acid type was used, one for each node. This battery can be replaced by a cap-lamp battery used in underground mines in compliance with Directorate General of Mine Safety-India (DGMS) standard. A Liquid Crystal Display (LCD) is programmed and connected to the microcontroller unit present at the receiver to display the desired output. The used transmitter and receiver units are shown in Figure 4.

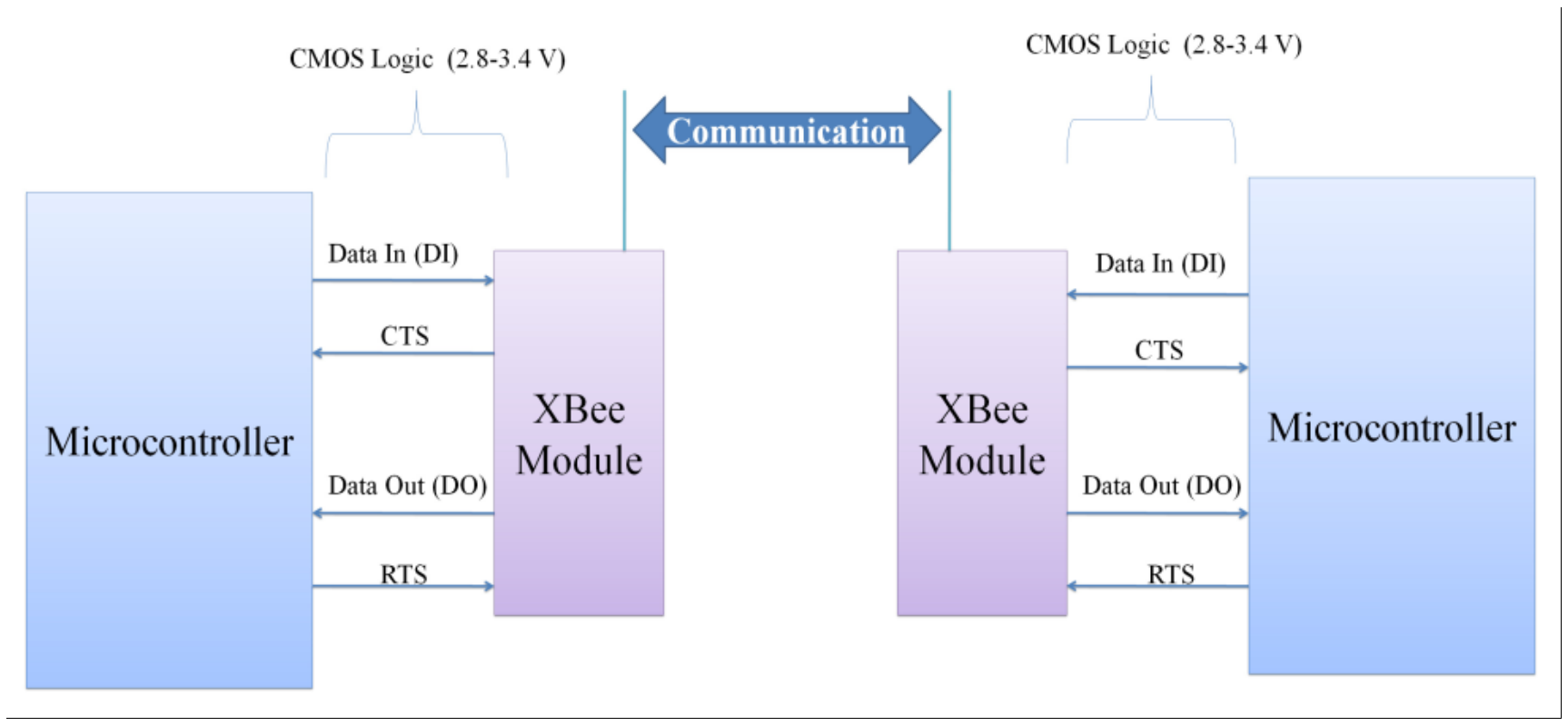

Fig.3: UART connections for the transmitter and receiver module

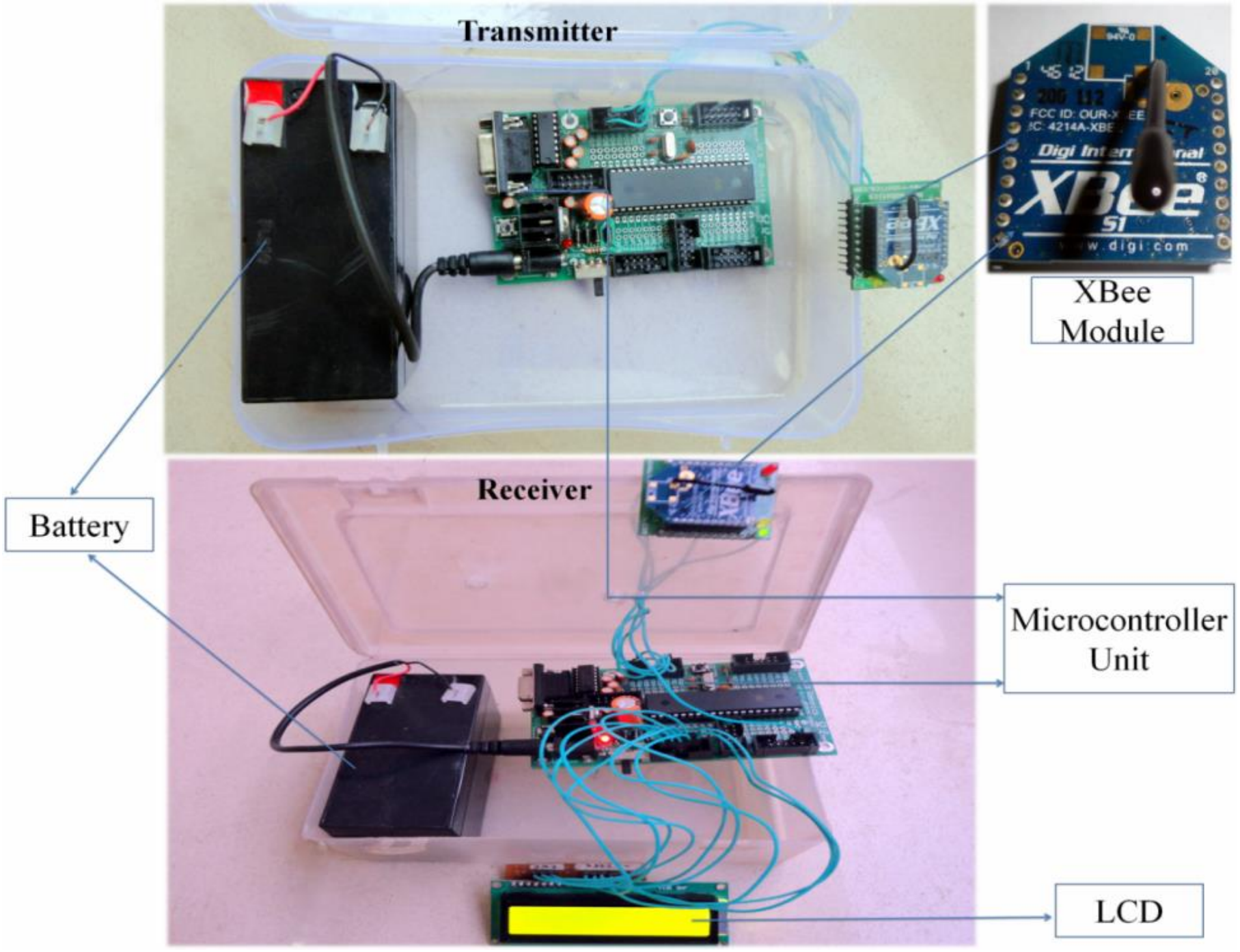

Fig.4: Transmitter and Receiver unit

For use in underground mine, the electronic instrument must be intrinsically safe to avoid any fire hazard. Since ZigBee protocol based wireless modules have been used in underground mines worldwide, they can be considered as intrinsically safe for most of the underground mining scenarios in India [19]. Parameters required for the XBee module to be intrinsically safe are specified in Table 2 [20]. The ZigBee protocol is based on the Carrier Sensing Multiple Access (CSMA) with Collision Avoidance (CA) channel access to provide energy saving, latency and negligible error in the received data packet. Direct Sequence Spread Spectrum (DSSS) modulation is used in the PHY layer that has high resistance for noise or jamming. ZigBee standard supports star, tree and mesh network, thus permitting numerous applications. In sleep mode it uses only $0.1\mu A$ that helps in energy saving during its idle period. It supports AES-128 encryption that converts a 128 bit plain-text to a 128 bit cipher-text. It has a capacity to acquire more than 256 peer to peer connections in a master-slave configuration; which is very high compared to other wireless protocols used in day to day life.

Table 2. Parameters required for intrinsically safe instrument

| **XBee Series 1 IEEE 802.1.5.4 Properties** | **Values** |
|---|---|
| Maximum power at antenna connector | 2mW |
| Maximum current at antenna connector | 7mA (AC current at 2.4GHz) |
| Sum total of all capacitance on PCB | 757pF |
| Sum total of all inductance on PCB | 60nH |
| Largest capacitor on PCB | 220pF |
| Largest inductor on PCB | 56nH |

The experiment was divided into two parts namely RSSI and Range test; RSSI test provides the data for determining path loss index and various parameters affecting the localisation and fading of power and Range test gives the operation range of aforesaid module in different underground mine scenarios.

*4.1.1. RSSI-test*

First set of readings were taken at the Longwall-face with shearer, hydraulic power supports, AFC, Stage Loader and other machineries which obstructed the wireless signal. To avoid fast fading of signal the readings were taken in a static environment free from the presence of moving machineries or men in between the transmitter-receiver pair. Second set of readings were taken beside the belt conveyor system, in running condition, installed in the gate-road of the mine, which would have created some fast fading.

#### *4.1.2. Range-test*

The range test was conducted in three different places, i.e. near the Longwall-face, the belt-conveyor system and in the inclined mine-car pathway.

### ***4.2. Experimental Procedure***

Firstly, RSSI-test was performed and readings were taken by fixing the transmitter node at the beginning of the Longwall-face close to the hydraulic powered roof support at a height of 1.5m from the floor. Both the transmitter and receiver setups were kept at a distance of 1m and 2m from the chocks and the working face respectively. The transmitter node was programmed to send 100 packets with a delay interval of 500ms between two subsequent packets and LCD showed the average RSSI over these 100 packets. Twenty number of RSSI readings were taken at each position of the receiver node and the same procedure was repeated up-to a distance of 20 meter with one meter step size. The Packet Received Rate (PRR) was also calculated and displayed on the LCD at one meter distance interval and all the readings were taken in line of sight condition. The second set of readings was taken on gate-road near the belt-conveyor system. The transmitter node was fixed at a location exactly 1m above the floor, half a metre away from the belt conveyor and the receiver node was kept at varying distances (1-20m) from the transmitter node along the passage.

The Range-test for the XBee module was then carried out sequentially in all the three areas by fixing the transmitter node at a particular location and moving the receiver node away from the LCD till it showed a ‘0’ value for the RSSI and indicated that, the packet sent by the transmitter could not be received beyond that particular distance.

## 5. Results and Analysis

The data collected near the working face and the belt-conveyor gate-road are represented in Table 3 and 4 respectively and the standard deviation was calculated for each set of RSSI values on every location.

The Standard Deviation ($SD$) can be calculated as

$$SD_i = \sqrt{\frac{\sum_{j=1}^{n}(X_j - M)^2}{n-1}} \tag{15}$$

Where, $SD_i$ is same as $Y_i$ of equation (9) for a particular distance $d_i$. $X_j$ represents the different RSSI values recorded at each distance $d_i$, $M$ is the mean RSSI and $n$ is the total number of observations i.e. 20. The integer variables $i$ and $j$ both vary from 1 to 20.

Table 3. Data collected near the longwall-face of GDK 10A

| Distance (m) | M (dBm) | SD (dBm) | PRR (%) |
|---|---|---|---|
| 1 | -51.65 | 0.48936 | 100 |
| 2 | -57.65 | 2.00722 | 100 |
| 3 | -71.5 | 4.54799 | 96.59 |
| 4 | -69.8 | 3.67924 | 96.76 |
| 5 | -73.95 | 5.78996 | 96.29 |
| 6 | -76.1 | 4.93004 | 95.83 |
| 7 | -76.85 | 5.83343 | 95.7 |
| 8 | -78.45 | 6.88665 | 95.07 |
| 9 | -80.25 | 6.04261 | 95.08 |
| 10 | -76.55 | 6.60522 | 95.45 |
| 11 | -76.8 | 5.94491 | 95.65 |
| 12 | -81.15 | 4.56828 | 93.92 |
| 13 | -80.95 | 3.64872 | 93.89 |
| 14 | -81.85 | 4.22119 | 93.9 |
| 15 | -79.35 | 3.54334 | 94.2 |
| 16 | -80.95 | 4.20443 | 93.77 |
| 17 | -82.6 | 4.87097 | 92.71 |
| 18 | -81.6 | 3.93901 | 93.85 |
| 19 | -84.15 | 4.51051 | 90.05 |
| 20 | -86.85 | 4.88041 | 86.2 |

Table 4. Data collected near the belt conveyor gate-road

| Distance (m) | M (dBm) | SD (dBm) | PRR (%) |
|---|---|---|---|
| 1 | -54.2857 | 3.48056 | 99.37 |
| 2 | -60.0952 | 1.92106 | 99.3 |
| 3 | -68.5714 | 7.59402 | 95.73 |
| 4 | -67.0476 | 7.89087 | 95.22 |
| 5 | -67 | 7.75887 | 96.19 |
| 6 | -73 | 4.12311 | 96.04 |
| 7 | -73.6667 | 6.5904 | 95.98 |
| 8 | -70.6191 | 5.45414 | 96.53 |
| 9 | -73.1905 | 6.14261 | 95.9 |
| 10 | -68.2381 | 5.76052 | 96.3 |
| 11 | -66.1905 | 4.44491 | 97.24 |
| 12 | -69.5714 | 3.35517 | 96.83 |
| 13 | -69 | 3.6606 | 96.89 |
| 14 | -75 | 5.12119 | 95.5 |
| 15 | -75.3333 | 4.23478 | 95.81 |
| 16 | -79.8095 | 4.7394 | 94 |
| 17 | -75.5714 | 3.99464 | 95.14 |
| 18 | -76.5714 | 5.59081 | 94.63 |
| 19 | -74.5455 | 5.41363 | 94.99 |
| 20 | -83 | 5.54076 | 92.8 |

MATLAB version 7.6.0.324 r2008a was used for the linear regression analysis model and the slope of the fitted gradient-line denotes the path loss index for the place of experiment, which was found to be 2.14. Figure 5 (A) depicts the scatter plot of received signal for Longwall-face corresponding to the logarithmic distance. It indicates that fading of signal was due to the presence of more number of obstructions. Moreover it also implies that more repeaters should be placed and the inter-node distance should be kept small as compared to typical outdoor scenario for which the index is 2. More fading and gradual degradation of power transmitted was due to the presence of metallic bodies; homogenous obstructions present in the surroundings and static nature of the environment offered less standard deviation (more concentrated in the region of 3.5 to 6) from the mean RSSI values. The values of PRR show a dependency on both standard deviation and received power with a higher correlation with the former. The signal is marginally affected for the first 3 to 4 metres by the waveguide property of the tunnel and the effect increases gradually afterwards. A trade-off is observed between distance covered and the wave guide effect, leading to a fluctuation of RSSI over a small range. As discussed in section 2, the curve fitting was done to find a relation between the standard deviation and distance to determine the coefficients for the Longwall mining area as shown in Figure 5 (B). The coefficients a, b, c, e and f of fourth degree polynomial are found to be $2.626 \times 10^{-6}$, $6.176 \times 10^{-3}$, -0.2276, 2.403 and -1.721 respectively. $R^2$ and Root Mean Square Error (RMSE) were also found out to be 0.8332 and 0.6958 respectively.

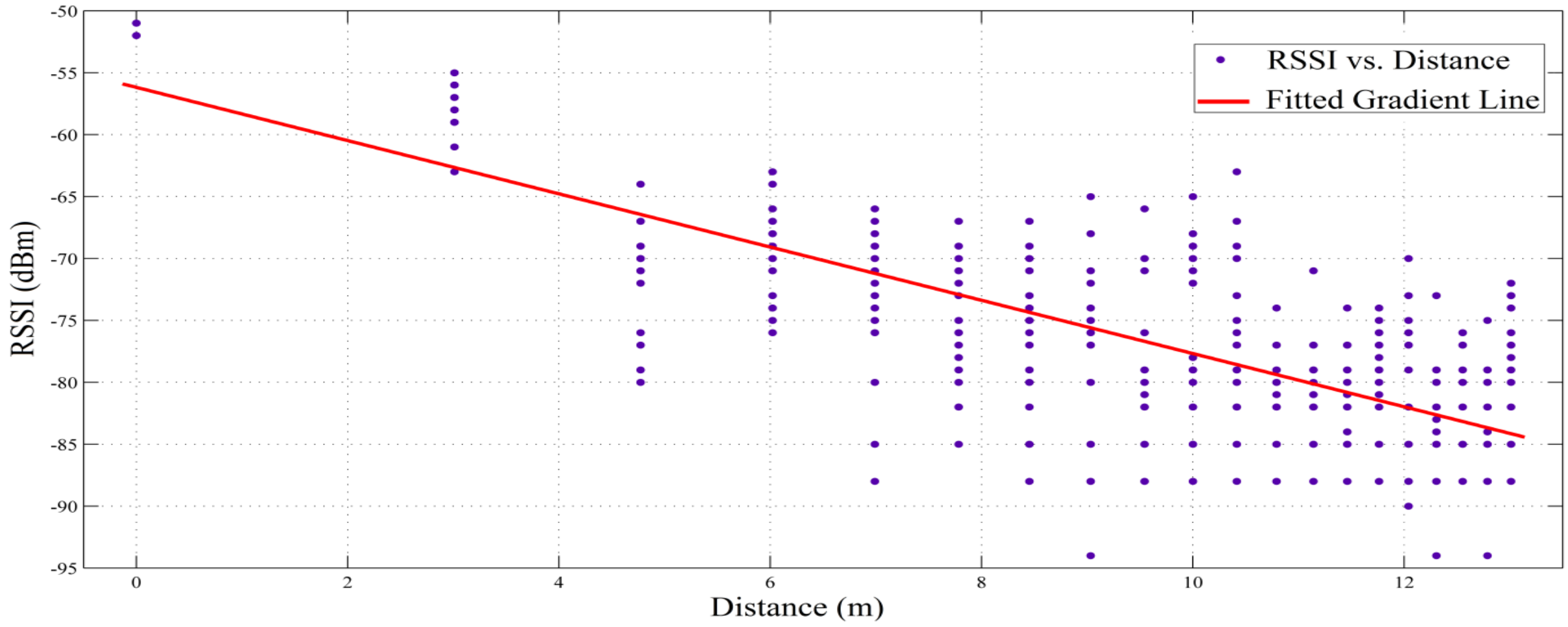

A)

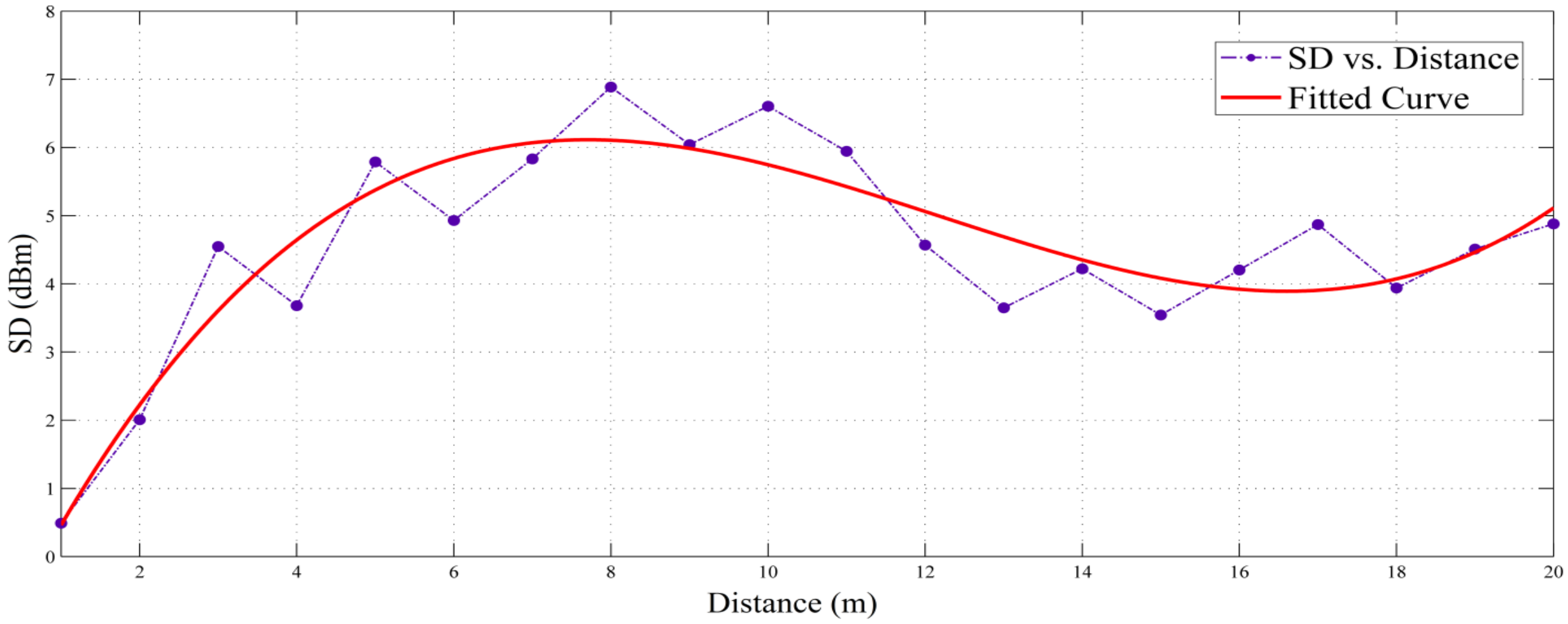

B)

Fig.5: A) Variation of RSSI with respect to distance near the Longwall-face. B) Relation between standard deviation and distance near the Longwall-face

For the belt-conveyor gate-road, the path loss index was found to be 1.568, using linear regression analysis. Figure 6 (A) depicts the scatter plot of RSSI vs. the logarithmic distance. The lower value of power loss compared to the Longwall-face was due to predominant effect of the waveguide property of tunnel. The standard deviations (more concentrated in the region of 4 to 7.5) from the mean RSSI values were high as compared to the Longwall-face area due to presence of inhomogeneous surroundings like different support systems, material and spacing between them, machineries, variable coal lump size carried by the belt and other distributive obstructions. Due to movement of the belt conveyor carrying coal lumps with variable sizes, some fast fading was found, as indicated by the dispersal of data from the fitted line. The signal loss for a particular place was found to be more, as compared to its consecutive place readings, each taken at one meter distance, due to presence of girders over the receiver. Presence of less metallic bodies in the gate-road compared to the Longwall-face lessened the fading effects. The signal propagation was mildly affected by the steel structure because the nodes were located higher than the belt conveyor support structure. Figure 6 (B) depicts the curve fitting for the fourth degree polynomial and the coefficients for determining the standard deviation as a function of distance was found to be $-6.685 \times 10^{-4}$, $0.3418 \times 10^{-1}$, $-0.5813$, $3.599$ and $-0.4563$ for a, b, c, e and f respectively. $R^2$ value of $0.474$ and RMSE value of $1.281$ indicate the fluctuation of standard deviation due to fast fading.

From the range-test, it was found that the XBee module provides satisfactory results up to a range of 40-45m, 60-65m, and 75-85m for the Longwall-face, belt-conveyor gate-road and mine-car pathway respectively.

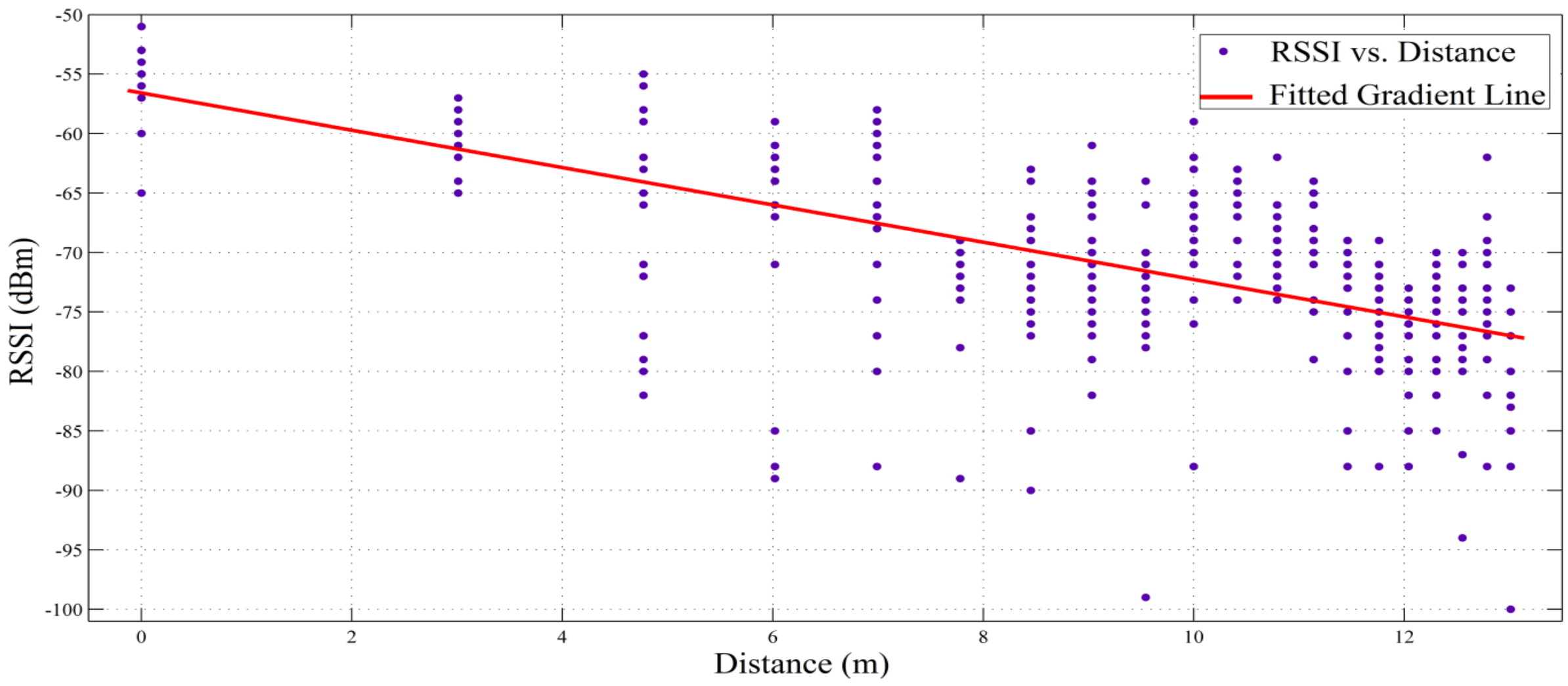

A)

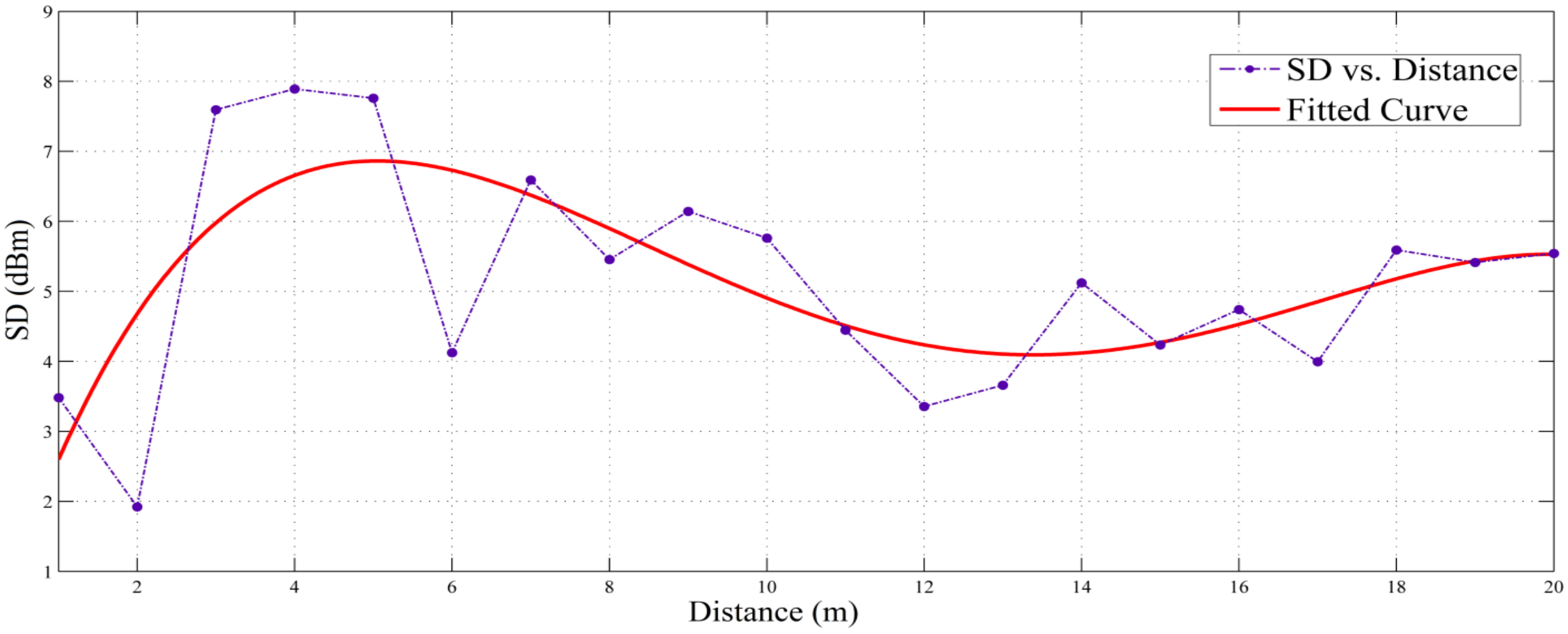

B)

Fig.6: A) Variation of RSSI with respect to distance in the belt conveyor gate-road. B) Relation between standard deviation and distance for the belt conveyor gate-road

## 6. Conclusion

This study reveals that the efficiency of a communication system is dependent on the underground mine-surroundings. Before implementing any wireless system in underground mines, the path loss index and the variance of Gaussian distribution representing the shadow fading effect for that place should be determined. This helps in determining the distance at which repeaters should be placed in order to enhance the signal and localise the sensor node from its received signal strength. With increasing number of physical obstructions, the value of path loss index increases, resulting in total loss of signal beyond a particular range. The XBee module facilitates satisfactory wireless communication over an adequate range of operation with negligible packet error rate. The PRR depends upon transmitter-distance and dynamic behaviour of the surroundings. These intrinsically-safe modules are economic, energy-efficient and fortify the mine safety system by enhancing tracking of miners and real-time data acquisition from sensors. The experiment was carried out in a hazard-prone underground coal mine. The experimental results may vary for different underground mines other than coal, due to the variation in earthy material, dimension of tunnels, passages, galleries and working areas depending on the mining method adopted. Two nodes have been used for experiment in this paper; to ensure viability of ZigBee protocol further study may be carried out to analyse the network performance using more than two nodes.

## Acknowledgement

We wish to express our sincere gratitude to the authorities of SCCL for permission and assistance in carrying out the experiment and collecting valuable data of GDK 10A. We thank the anonymous reviewers for their valuable comments.